\documentclass[aps,prb,amsmath,amssymb,twocolumn,superscriptaddress]{revtex4-2}

\usepackage{graphicx}
\usepackage{bm}
\usepackage{hyperref}
\usepackage{dsfont}
\usepackage{color}

\usepackage{ulem}
\usepackage{xcolor}
\usepackage{dcolumn}
\usepackage{wasysym} 
\usepackage{xspace}
\usepackage{ulem}

\newcommand{\ve}[1]{\boldsymbol{#1}}

\graphicspath{{./Figures}}

\begin{document}

\title{
	Alteration of Topology in Quantum Phase Transitions via Symmetry Enrichment
}
\author{Gabriel Rein}
\email{gabriel.rein@uni-wuerzburg.de}
\affiliation{\mbox{Institut f\"ur Theoretische Physik und Astrophysik,
		Universit\"at W\"urzburg, 97074 W\"urzburg, Germany}}
\affiliation{W\"urzburg-Dresden Cluster of Excellence ct.qmat, Am Hubland, 97074 W\"urzburg, Germany}
\author{Marcin Raczkowski}%
\affiliation{\mbox{Institut f\"ur Theoretische Physik und Astrophysik,
		Universit\"at W\"urzburg, 97074 W\"urzburg, Germany}}
\author{Zhenjiu Wang}%
\affiliation{\mbox{Ludwig-Maximilians-Universit\"at M\"unchen, Theresienstr. 37, 80333 M\"unchen, Germany}}
\author{Toshihiro Sato}%
\affiliation{W\"urzburg-Dresden Cluster of Excellence ct.qmat, Am Hubland, 97074 W\"urzburg, Germany}
\affiliation{\mbox{Institute for Theoretical Solid Sate Physics, IFW Dresden, Helmholtzstr. 20, 01069 Dresden, Germany}}
\author{Fakher F. Assaad}%
\affiliation{\mbox{Institut f\"ur Theoretische Physik und Astrophysik,
		Universit\"at W\"urzburg, 97074 W\"urzburg, Germany}}
\affiliation{W\"urzburg-Dresden Cluster of Excellence ct.qmat, Am Hubland, 97074 W\"urzburg, Germany}


\date{\today}

\begin{abstract}
Topology  plays a cardinal role in explaining phases and quantum phase transitions beyond the Landau-Ginzburg-Wilson paradigm.  In this study, we formulate a set of models of Dirac fermions in 2+1 dimensions with SU($N$)$\times$SU(2)$\times$U(1) symmetry that have the potential to host critical points described by  field theories  with topological terms. For $N=2$ it shows a rich phase diagram containing semimetallic, quantum spin Hall insulating, Kekulé valence bond solid and s-wave superconducting phases and features multiple Landau-Ginzburg-Wilson phase transitions driven by interaction strength. At $N=1$ a deconfined quantum critical point is observed. At $N=2$ one expects the critical 
theory to correspond to  a level 2 Wess-Zumino-Witten  theory in 2+1 dimensions. Here the numerical results however show a strong first order transition.  Another transition can be governed by a topological $\theta$-term which is rendered irrelevant for even values of $N$  thus leading to Landau-Ginzburg-Wilson behaviour. 
\end{abstract}
\maketitle


\paragraph*{Introduction.\textemdash\label{sec:Introduction}}
Topology plays a pivotal role in the solid state: It provides an elegant understanding 
of phases and phase transitions beyond the Landau-Ginzburg-Wilson paradigm 
\cite{Kosterlitz73,Haldane1983,Senthil-PhysRevB2004,Moore-QSHE,Balents10}. 
Specific examples that set the stage for this Letter include antiferromagnetic 
spin-$S$ chains, with low energy fluctuations described by an  1+1 dimensional SO(3) 
non-linear sigma model with topological $\theta$-terms at $\theta =  2 \pi S $. This provides a 
topological understanding of the distinct physics of half-integer and integer spin chains 
\cite{ Haldane1983,Dagotto96}. By  tuning the strength of the second nearest neighbor 
antiferromagnetic exchange coupling, the spin-1/2 chain can undergo a continuous phase transition to a valence  bond solid (VBS) phase that breaks site-inversion symmetry \cite{Majumdar69,Eggert96}.  This transition is described by a level $k=1$ Wess-Zumino-Witten (WZW) theory with emergent SO(4) symmetry that locks in spin-spin  and dimer-dimer 
correlations.  The topological term accounts for the non-trivial structure of the VBS domain wall  that hosts a   spin-1/2 degree of freedom. The phase transition out of the VBS state occurs via the proliferation of domain walls and concomitant appearance of critical antiferromagnetic fluctuations. Importantly, the $k=1$ Wess-Zumino-Witten  theory has a marginal SO(4) symmetry breaking operator the relevance of which depends upon its sign \cite{Haldane-WZW,Senthil24}. Thereby the  transition requires no multiparameter fine tuning.  Higher half-integer  spin chains map onto  $k=2S$ Wess-Zumino-Witten  theories that correspond to multi-critical points such that fine tuning  is necessary to study them \cite{Haldane-WZW,Chepiga-WZW}. 

The generalization of the above to 2+1 dimensions has broadly lead to the theory of deconfined quantum criticality (DQC) \cite{Haldane88,READ1989609,Schwab23,Senthil-Science,Senthil-PhysRevB2004,Senthil-Fisher-theta-term,Tanaka2005}.  The physics of the   easy-plane  DQC point (DQCP) as well as the criticality of the  model presented in  Ref.~\cite{Sato-Assaad-2017} can be understood in terms of a 2+1 dimensional  SO(4)  nonlinear sigma model  with $\theta$-term at $\theta= \pi$ \cite{Senthil-Fisher-theta-term,Abanov2000}.  On the other hand the  DQCP, as realized in 
the $J-Q$ \cite{Sandvik07}, loop \cite{Nahum15,Nahum15_1}  or  various  fermionic models \cite{Yuhai-Skyrmions,Goetz23},   can map onto the 2+1  dimensional   SO(5)  nonlinear model  with level $k=1$ WZW  term \cite{Lee-Sachdev-2015,Assaad-PhysRevLett.126.045701}.  There is an  ongoing discussion  pertaining  to the nature,  weakly first order or  continuous,  of  the aforementioned  critical points \cite{Troyer-DQCP,WangC17,Nahum19,Takahashi24}. For our purposes, the mere proximity  of the models to a unitary critical point where topology, in terms of  $\theta$-terms  at  $\theta = \pi$  or level $k=1$ WZW terms is of  importance. 
The aim of this Letter is to propose a symmetry enhancement  of  the model of Ref.~\cite{Yuhai-Skyrmions} so as to modify  the topological aspects  of the  putative  field theory to level $k=N$ WZW terms and $\theta$-terms  at  $\theta = N\pi$.  Such a  class of models,  allows us to switch on and off the role of topology, and thereby between  Landau-Ginzburg-Wilson and topological quantum phase transitions.  

\begin{figure}
	\centering
	\includegraphics[width=0.95\columnwidth]{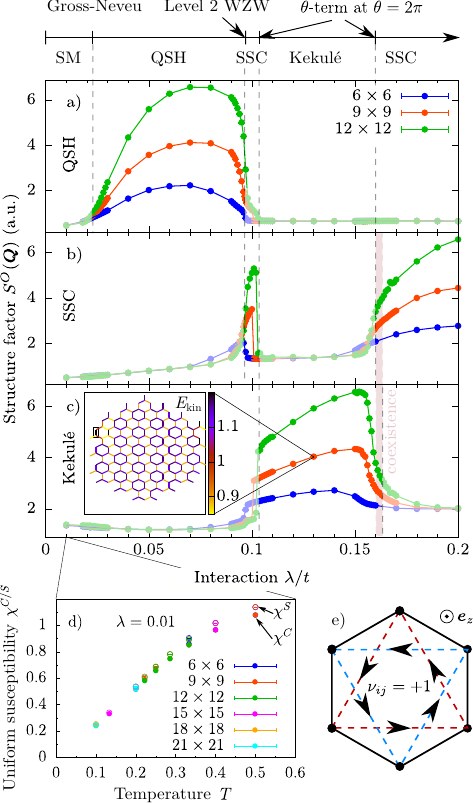}
	\caption{Reconstruction of the phase diagram of the model $\hat{H}_t+\hat{H}_\lambda$ in Eqs. (\ref{eq:Model-hopping}),(\ref{eq:Model-interaction}) for the case $N=2$. The structure factors for (a) QSH order at the $\Gamma$-point, (b) SSC order at the $\Gamma$-point and (c) Kekulé order at the Dirac-point versus $\lambda$ obtained by tracing over the orbitals in Eq. (\ref{eq:structurefactor}) are employed to draw the phase diagram at the top. The inset in (c) shows the kinetic energy per bond in real space at $\lambda=0.13$ upon pinning the hopping to the rimmed bond. (d) Uniform charge (filled) and spin (unfilled symbols) susceptibilty at $\lambda=0.01$. (e) Hexagon illustrating the interaction term in Eq. (\ref{eq:Model-interaction}). For the depicted direction of bonds $\nu_{\boldsymbol{i}\boldsymbol{j}}=+1$. The error bars are smaller than the symbol size for most of the data.}
	\label{fig:structurefactor}
\end{figure}

\paragraph*{Model.\textemdash\label{sec:Model}}
The model we consider comprises a nearest-neighbor hopping term on a honeycomb lattice
\begin{equation}
	\hat{H}_t=-t\sum_{\langle\boldsymbol{i},\boldsymbol{j}\rangle}\sum_{s=1}^{N}\big(\boldsymbol{c}_{\boldsymbol{i},s}^\dagger\boldsymbol{c}_{\boldsymbol{j},s}+\textrm{h.c.}\big)
	\label{eq:Model-hopping}
\end{equation}
supplemented with an interaction term characterized by the square of the Kane-Mele term
\begin{equation}
	\hat{H}_{\lambda}=-\frac{\lambda}{N}\sum_{\varhexagon}\Big(\sum_{s=1}^{N}\sum_{\langle\langle\boldsymbol{i},\boldsymbol{j}\rangle\rangle\in \varhexagon}i\nu_{\boldsymbol{i}\boldsymbol{j}}\boldsymbol{c}_{\boldsymbol{i},s}^\dagger\boldsymbol{\sigma}\boldsymbol{c}_{\boldsymbol{j},s}+\textrm{h.c.}\Big)^2
	\label{eq:Model-interaction}
\end{equation}
where $t$ is a hopping parameter and $\lambda$ is an interaction strength. $\boldsymbol{c}_{\boldsymbol{i},s}^\dagger=\big(c_{\boldsymbol{i},s,\uparrow}^\dagger,c_{\boldsymbol{i},s,\downarrow}^\dagger\big)$ is a spinor containing fermionic operators that create an electron on lattice site $\boldsymbol{i}$ with $z$-component of spin $\sigma$ and carry flavor $s$. The hopping occurs only on nearest neighbor bonds $\langle\boldsymbol{i},\boldsymbol{j}\rangle$ whereas the interaction involves next-nearest neighbor pairs $\langle\langle\boldsymbol{i},\boldsymbol{j}\rangle\rangle$ that are related to phase factors $\nu_{\boldsymbol{i}\boldsymbol{j}}=\pm 1$ defining a direction as in the canonical Kane-Mele model \cite{Kane-Mele-model}. This is exemplified in Fig. \ref{fig:structurefactor}(e). The vector $\boldsymbol{\sigma}$ contains the Pauli spin matrices. The model preserves SU(2) spin symmetry as well as U(1) charge and SU($N$) flavor symmetry.
The ground state phase diagram at $N=1$ has been studied in Ref.~\cite{Yuhai-Skyrmions,Zhenjiu-QSH-SC}. As a function of the interaction strength, $\lambda$,   one observes a Gross-Neveu O(3) transition from a Dirac semimetal (SM) to 
a dynamically   generated quantum spin Hall (QSH)  phase \cite{Liu21}.   Upon  further increasing $\lambda$  the QSH state  gives way  to an s-wave superconducting (SSC) state. This phase  transition  is an instance  of  a  DQCP  with  emergent SO(5) symmetry \cite{Toshihiro-mass-terms}.  It is driven by condensation of skyrmions of the SO(3) QSH
order parameter  that carry charge $2e$ \cite{Grover-Senthil-2008} and that condense at the critical point. 

We study the model defined by $\hat{H}=\hat{H}_t+\hat{H}_{\lambda}$ by employing a finite temperature  auxiliary field Quantum Monte Carlo (QMC) algorithm \cite{QMC-PhysRevD.24.2278,QMC-PhysRevB.40.506,Assaad2008-QMC} based on the ALF package \cite{ALF}.   In this approach, the perfect square term is  decoupled via a Hubbard-Stratonovich transformation with a space ($\ve{i}$) and time ($\tau$) auxiliary field $\ve{\Phi}(\ve{i},\tau)$ and  the partition function reads
\begin{equation}
\label{Eq:QMC}
	Z =  \int D \left\{ \ve{\Phi}(\ve{i},\tau) \right\} e^{- N S\left\{ \ve{\Phi}(\ve{i},\tau) \right\}}.
\end{equation}
Owing to time reversal symmetry, the action is real for  $\lambda>0$ \cite{Wu04} such that  $e^{- N S\left\{ \ve{\Phi}(\ve{i},\tau) \right\}}$ can be used as a  weight function for  Monte Carlo importance sampling. Since the Hubbard-Stratonovich transformation does not break the SU($N$) flavour symmetry,  simulations of the higher symmetry 
model merely require multiplying the action by $N$ \cite{Assaad04}.  As we will argue in the discussion section, the form  of the action equally  suggests that if the low energy physics at $N=1$ is dominated by  a level $k=1$  WZW term then the corresponding critical point in the  SU($N$) model is described by a level $k=N$ WZW term \cite{Lee-Sachdev-2015}. 
For all of the presented calculations we set $t=1$ and consider a half-filled band, i.e. one electron per site corresponding to a vanishing chemical potential. Moreover, we use a symmetric Trotter decomposition with an imaginary time discretization $\Delta\tau=0.1$ and employ a finite temperature code at $\beta=2L$. Looking ahead this choice is justified, since the symmetry breaking phases  as well as  the SM we encounter all 
have a dynamical exponent set by unity. The imaginary-time discretization $\Delta\tau$ leads to a systematic error that scales as $\Delta\tau^2$. 
This error must remain well below characteristic low-energy scales of the system to avoid contaminating physical observables. 
In the Supplemental Material of Ref. \cite{Yuhai-Skyrmions}, it was shown that for $N=1$, choosing $\Delta\tau=0.2$ keeps such errors negligible.
Meanwhile, Ref. \cite{Schwab23} demonstrated that for larger $N$, setting $\Delta\tau \propto 1/N$ is a suitable guideline to ensure these errors remain below the relevant energy scale.
Based on these results, we choose the aforementioned value of $\Delta\tau$ in our simulations.

\paragraph*{Results.\textemdash\label{sec:Results}}
Our main results are summarized in the phase diagram of the model as shown in Fig \ref{fig:structurefactor}. Here, we study the phase diagram for $N=2$ by calculating structure factors
\begin{eqnarray}
	S_{\delta,\delta'}^{O}(\boldsymbol{q})=\frac{1}{L^2}\sum_{\boldsymbol{r},\boldsymbol{r}'}e^{i\boldsymbol{q}\cdot(\boldsymbol{r}-\boldsymbol{r}')}\big\langle\hat{\boldsymbol{O}}_{\boldsymbol{r},\delta}\hat{\boldsymbol{O}}_{\boldsymbol{r}',\delta'}\big\rangle
	\label{eq:structurefactor}
\end{eqnarray}
for different observables $O$ defined by the order parameter $\hat{\boldsymbol{O}}_{\boldsymbol{r},\delta}$, where $\boldsymbol{r}$ is the position of a unit cell, $\delta$ denotes an orbital and $L$ is the lattice size. $\big\langle\hat{\boldsymbol{O}}_{\boldsymbol{r},\delta}\hat{\boldsymbol{O}}_{\boldsymbol{r}',\delta'}\big\rangle$ is a cumulant as the background remains to be subtracted yet vanishing for most observables. The QSH order parameter can be identified as a spin-orbit coupling $\hat{\boldsymbol{O}}_{\boldsymbol{r},\delta}^{\rm{QSH}}=\sum_{s=1}^N i\nu_{\boldsymbol{i}_{\delta}\boldsymbol{j}_{\delta}} \boldsymbol{c}_{\boldsymbol{r}+\boldsymbol{i}_{\delta},s}^\dagger \boldsymbol{\sigma} \boldsymbol{c}_{\boldsymbol{r}+\boldsymbol{j}_{\delta},s}+\rm{h.c.}$
where $\boldsymbol{i}_{\delta}$ and $\boldsymbol{j}_{\delta}$ are the legs of next-nearest neighbor bonds $\delta =1,\ldots,6$ inside a hexagon. This operator is closely related to the interaction Hamiltonian in Eq. (\ref{eq:Model-interaction}). As the SSC order parameter we define the s-wave pairing operator 
$\hat{O}_{\boldsymbol{r},\delta}^{\rm{SSC}}=\sum_{s=1}^N c_{\boldsymbol{r},\delta,s,\uparrow}^\dagger c_{\boldsymbol{r},\delta,s,\downarrow}^\dagger + \rm{h.c.}$
where $\delta$ now reflects the physical orbital of unit cell $\boldsymbol{r}$. Eventually, for the order parameter resolving Kekulé VBS order we choose the kinetic energy of a single bond $\hat{O}_{\boldsymbol{r},\delta}^{\rm{VBS}}=\sum_{s=1}^N\sum_{\sigma=\uparrow,\downarrow}c_{\boldsymbol{r}+\boldsymbol{i}_{\delta},s,\sigma}^\dagger c_{\boldsymbol{r}+\boldsymbol{j}_{\delta},s,\sigma} + \rm{h.c.}$
where $\boldsymbol{i}_{\delta}$ and $\boldsymbol{j}_{\delta}$ are the legs of nearest neighbor bonds $\delta=1,2,3$ in the honeycomb lattice.

Similar to the $N=1$ case we find a SM phase for small values of $\lambda$ as can be proposed out of Fig. \ref{fig:structurefactor}(d) which shows the defining linear dependence of the uniform charge and spin susceptibility $\chi^{C/S}(\boldsymbol{q}=\boldsymbol{0})$ on temperature, approaching zero as temperature declines. As spin and charge are conserved in the Hamiltonian the uniform susceptibility is given by $\chi^O(\boldsymbol{0})=\beta\sum_{\delta,\delta'}S_{\delta,\delta'}^O(\boldsymbol{0})$.
The QSH phase can be identified by considering the divergent behavior of the structure factor with respect to the system size as shown in Fig. \ref{fig:structurefactor}(a). This can be supported by a vanishing $\chi^C$ indicating a charge gap alongside gapless spin excitations specified by a finite $\chi^S$. Data for $\chi^{C/S}$ is shown in the Supplemental Material \cite{supplemental}. The diverging structure factor of SSC correlations appearing for greater values of $\lambda$ viewed in Fig. \ref{fig:structurefactor}(b) together with a finite $\chi^C$ reveal an SSC phase.

In contrast to the $N=1$ case, we observe the emergence of a Kekulé VBS phase that suppresses the SSC phase partially. This manifests again in the structure factor of Fig. \ref{fig:structurefactor}(c) as well as a vanishing $\chi^C$ and $\chi^S$ indicating a trivial insulator. Additionally, the inset in Fig. \ref{fig:structurefactor}(c) reveals the Kekulé pattern recovered by plotting the kinetic energy per bond upon introducing a pinning field to the hopping breaking translational symmetry. It can be obtained by calculating the expectation value of the VBS order parameter.

A common measure for studying the nature of phase transitions is the renormalization-group invariant correlation ratio
\begin{equation}
	R^O
	= 1 - \frac{S^O(\boldsymbol{Q}+\Delta\boldsymbol{q})}{S^O(\boldsymbol{Q})}
\end{equation}
with $|\Delta\boldsymbol{q}|=\frac{4\pi}{\sqrt{3}L}$ and ordering wave vector $\boldsymbol{Q}$ which is zero for the considered QSH and SSC correlations 
\cite{Binder-ratio,Pujari-PhysRevLett.117.086404}. In our simulations we use a scaling of $\beta=2L$ assuming a dynamical critical exponent $z=1$ which corresponds to a linear behaviour of the susceptibilities with respect to temperature \cite{Hohenadler-structurefactor}.\\
\begin{figure}
	\centering
	\includegraphics[width=\columnwidth]{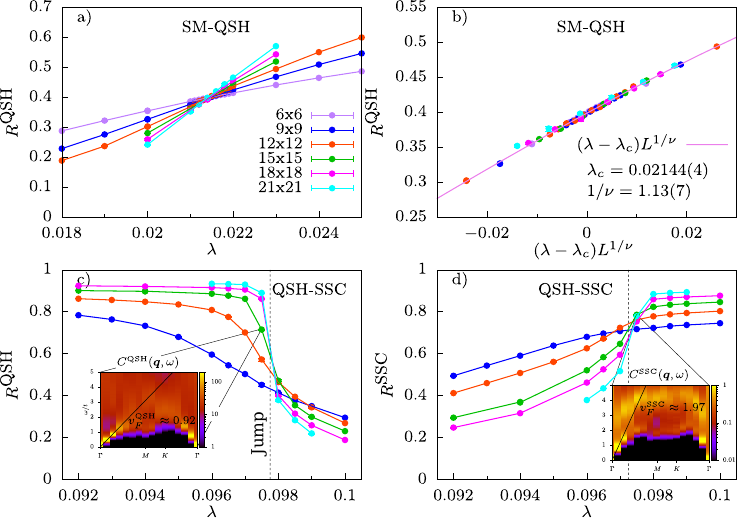}
	\caption{Correlation ratio of QSH order for the SM-QSH transition (a) together with a data collapse (b) employed to extract the values of $\lambda_c$ and $1/\nu$ by fitting to $L^{1/\nu}(\lambda-\lambda_c)$. Correlation ratio of QSH (c) and SSC (d) order parameters. The inset shows dynamical structure factors for $L=15$ at the QSH-SSC transition $\lambda=0.0975$. The slope denotes the respective Fermi velocity extracted from the Dirac cones in units of hopping amplitude $t$ times lattice constant $a$.}
	\label{fig:SM-QSH-SSC}
\end{figure}
For the SM-QSH transition the correlation ratio is shown in Fig. \ref{fig:SM-QSH-SSC}(a) together with a data collapse \ref{fig:SM-QSH-SSC}(b) embodying a fit of the data to $L^{1/\nu}(\lambda-\lambda_c)$, which reveals a transition point at $\lambda_c=0.02144(4)$ and an inverse correlation length exponent $1/\nu=1.13(7)$. 
An analogue analysis for the correlation ratios of QSH and SSC order around the QSH-SSC transition as presented in Figs. \ref{fig:SM-QSH-SSC}(c) and \ref{fig:SM-QSH-SSC}(d) reveals a jump for the larger lattices. These data points do not collapse nicely for any choice of parameters $\lambda_c,\nu$. The insets show dynamical structure factors obtained from time-displaced correlation functions as extracted with the ALF \cite{ALF} implementation of the stochastic analytical continuation approach \cite{Sandvik-PhysRevB.57.10287,beach2004}. The velocities in the particle-hole and particle-particle channels can be extracted from the slope of the spectral functions. It turns out that at the presumed transition point $\lambda=0.0975$ the velocities of QSH and SSC order parameters differ by a factor of around 2 such that the numerical results do not support emergent Lorentz invariance. Otherwise, we would observe only one velocity for all modes.

\begin{figure}
	\centering
	\includegraphics[width=\columnwidth]{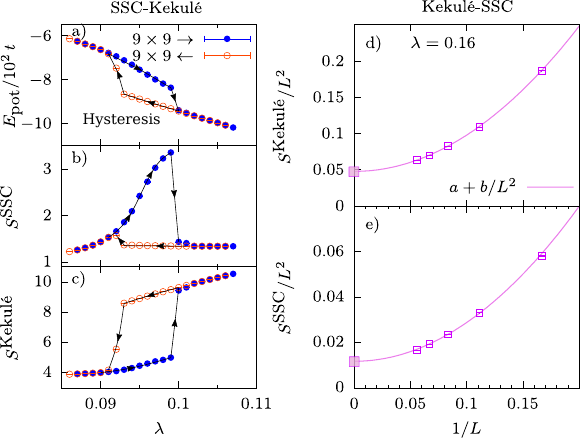}
	\caption{Hysteresis around the SSC-Kekulé transition on the $9\times9$ lattice for (a) the potential energy and the structure factor for (b) SSC order and (c) Kekulé order. The curves for increasing and decreasing $\lambda$ are distinguished by color. Extrapolation of the scaled structure factors (d) $S^{\textrm{Kekulé}}/L^2$ and (e) $S^{\rm{SSC}}/L^2$ to $1/L\to 0$ for $\lambda=0.16$ by fitting the data to a second order polynomial.}
	\label{fig:SSC-Kekule-SSC}
\end{figure}

The narrow SSC phase around $\lambda=0.1$ undergoes another transition to a Kekulé-VBS which is marked by a clear jump in the structure factors of the involved order parameters already on small lattices, as shown in Figs. \ref{fig:structurefactor}(b) and \ref{fig:structurefactor}(c). To further analyse the behavior of this jump, we plot the observables by using the QMC configuration of a given $\lambda$ as input for the simulation at the next step $\lambda\pm\Delta\lambda$. This results in the hysteresis curves shown in Figs. \ref{fig:SSC-Kekule-SSC}(a)-(c) for the potential energy as well as the structure factors of SSC and Kekulé correlations. Notably, the transition appears to be highly sensitive with respect to the lattice size, which, in combination with exceptionally long warm-ups, complicates determining the exact transition point. This sensitivity also impairs the analysis of the QSH-SSC transition since the transitions overlap for small lattice sizes.

Remarkably, the next transition from Kekulé to SSC features a coexistence regime as evident from Figs. \ref{fig:SSC-Kekule-SSC}(d) and \ref{fig:SSC-Kekule-SSC}(e), where we extrapolate the scaled structure factors of Kekulé and SSC correlations to $L\to\infty$ by means of a second order polynomial. This reveals that both Kekulé and SSC order are present as further detailed in the Supplemental Material \cite{supplemental}.

\paragraph*{Discussion.\textemdash}
In comparison to the $N=1$ phase diagram the model at $N=2$  shows an additional Kekulé VBS phase.  It is interesting to view this  result in the $N$-$\lambda$ plane.  In the large-$N$ limit, owing to Eq.~(\ref{Eq:QMC}), one expects  the saddle 
point approximation to be exact. As shown in the supplemental material \cite{supplemental} 
the SM phase at small $\lambda$ gives way to the QSH phase with mean-field order parameter exponent $\beta = 1$. At  finite values of $N$, one can account for the  SSC phase  by expanding the perfect square term, thereby revealing the presence of an attractive interaction in the s-wave channel \cite{Zhenjiu-QSH-SC}.  Another way of understanding the SSC phase, is  via the use of a Fierz identity (see Ref. \cite{Igor-Fierz-identity} and Supplemental Material \cite{supplemental}).  In particular since the QSH and SSC order parameters anti-commute the  QSH  mass term naturally also allows for an SSC instability. The Kekulé phase is surrounded by the SSC phases. The Kekulé and SSC mass terms are compatible (i.e. anticommute) such that one order potentially triggers the other according to Fierz identity. In the Supplemental Material \cite{supplemental} we describe, that unlike other VBS phases observed in the literature following Ref. \cite{Read-Sachdev-PhysRevLett.62.1694} the VBS observed here has no mean-field counterpart in the model. Here, the VBS vanishes together with the SSC as $N\to\infty$. Importantly,  the transitions we observe between ordered phases are 
between compatible (i.e. anticommuting) mass terms that in principle invalidate a Landau-Ginzburg-Wilson description:  In $d+1$-dimensions,  $d+3$ ($d+2$)  compatible mass terms lead  to WZW ($\theta$)- terms \cite{Abanov2000}.

We now discuss the nature of the transitions. The SM to QSH  transition falls in the category of 
Gross-Neveu criticality \cite{Gross74,Herbut09} and 
corresponds to a Landau-Ginzburg-Wilson transition where topology does not play a role.  One expects the transition to remain continuous as a function of $N$ albeit with $N$-dependent exponents. This stands in agreement with  our  numerical results.

We observe two transitions between the Kekulé and SSC phases. One is a first order transition, and the second shows a coexistence regime which can be argued to be called a supersolid \cite{Onsager-supersolid,Andreev-1971,Chester-PhysRevA.2.256,Legget-PhysRevLett.25.1543,Batrouni-PhysRevLett.74.2527,Batrouni-PhysRevLett.84.1599,Fisher-supersolid,Fisher-PhysRevLett.32.1350,Frey-PhysRevB.49.9723}, whose thermodynamic stability can be confirmed by a finite uniform charge susceptibility as shown in the Supplemental Material \cite{supplemental}. These are typical for Landau-Ginzburg-Wilson  transitions between  different symmetry broken phases, $U(1)$ charge and  $U(1)$ rotational symmetry in the continuum. However, in our setting this is surprising since the  two SSC, $M_1$, $M_2$ and two Kekulé, $M_3$, $M_4$ masses anticommute \cite{Mudry2009}. Let us start with the $N=1$ case, and  supplement the  Dirac vacuum with an interaction term $ -U\sum_{i=1}^{4} (M_i)^2 $. Since $\Gamma_{i,j} = \frac{i}{2} \left[ M_i,M_j\right] $  correspond  to the six generators of SO(4) the Hamiltonian possesses   SO(4) global  symmetry \footnote{A  rotation $ e^{i \theta \ve{e}_{\alpha} \Gamma_{\alpha} }\Psi(x) $ of  the Dirac spinor,   amounts to an SO(4) rotation of the vector of mass terms. }. 
In the massive phase, one can 
integrate out the fermions to obtain a $\theta$-term  at $\theta = \pi$ \cite{Abanov2000,Senthil-Fisher-theta-term} that accounts for the fact that the vortex of the U(1) Kekulé order parameter shows an emergent spin-1/2  degree of freedom.   
In Refs.~\cite{Sato-Assaad-2017,Toshihiro-Topology} such a transition with emergent SO(4) symmetry has been observed numerically and the consequences of the topological term were discussed in detail.  We  now have to account for the SU($N$) flavor symmetry. Importantly, this symmetry is not broken in our simulations (See supplemental material \cite{supplemental}). Under this condition, $N$ is merely a multiplicative factor of the action:  
\begin{align}
	&S_{\theta}^N=\rho_s\,N\,\int\text{d}^3x(\partial_{\mu}\boldsymbol{n})^2\nonumber\\
	&+iN\pi\underbrace{\frac{1}{\text{Area}(S^3)}\int\text{d}^3x\,\varepsilon^{\alpha\beta\gamma\delta}n_{\alpha}\partial_{x_1}n_{\beta}\partial_{x_2}n_{\gamma}\partial_{x_3}n_{\delta}}_{Q\in\mathbb{Z}}. 
\end{align} 
It is now apparent  that for even values of $N$  the topological term can be omitted such that the  observed phase transitions  are described by  Landau-Ginzburg-Wilson order parameter theories. 

The case of the QSH to SSC transition is very similar.  In fact the five masses build a quintuplet of anti-commuting mass terms \cite{Mudry2009}. At $N=1$, there is ample evidence that this phase transition is partially described by a level $k=1$ WZW theory with  emergent SO(5) symmetry. Repeating the same argument as above, and provided that the SU($N$) symmetry is not broken,  the relevant theory is  that of the level $k=N$ WZW model.   At $N=2$ our results show a first order transition as apparent from the jump in the correlation ratio. This is supported by the absence of emergent Lorentz symmetry preventing a description by means of a Lorentz invariant theory. A possible interpretation of this result is obtained by  analogy to WZW criticality in 1+1 dimensions:  WZW criticality in 2+1 dimensions describes a multi-critical point due to additional relevant operators \cite{Haldane-WZW,Takahashi24}.   Since our model has only a single parameter, it does not allow for fine tuning, and we generically expect a first-order transition. Another interpretation stems from the notion of pseudo-criticality already present at $N=1$ \cite{WangC17,WangC19,Nahum19,Hawashin23}:  the level of  the  WZW term enhances the  \textit{distance} from the non-unitary critical point  and ultimately renders the transition strongly first order.
We note  that transitions between valence bond solids and antiferromagnets for $S=1$  in 2 spatial dimensions  could correspond to a  level 2 WZW theory.  Numerically,  such transitions \cite{Wildboer20} turn out to be of first order.

\paragraph*{Conclusion.\textemdash}
Symmetry enhancement of models allow for rich phase diagrams and novel phases.  The specific model we  consider
is of great interest since it hosts phase transitions between compatible orders  of  Dirac systems.  This 
is known to define a route for generating quantum phase transitions where topology plays a pivotal role \cite{Abanov2000}. 
By adding  $N$ flavors to the fermions, our model  acquires an extra SU($N$) symmetry. Provided, that this symmetry is not spontaneously broken, topological terms required to understand the quantum critical point in the $N=1$ theory will be multiplied by $N$. Hence 
$\theta$-terms at $\theta = \pi N$  and level $k = N$ WZW theories can be generated in 2+1 dimensions. We have concentrated on the $N=2$ case, and  shown that in contrast to the $N=1$ case, all the transitions we observe lie in the realm of Landau-Ginzburg-Wilson transitions. For the $\theta$-term topology evidently 
drops out at $\theta =   2 \pi$.  As in the one-dimensional case, we understand that the level $k=2$ WZW theory corresponds to a multicritical point that  is hard to reach within our parameter space. 

Clearly, future work will include simulations at larger values of $N$. In this context, we note that  the honeycomb lattice regularization 
we have considered reduces the U(1) VBS symmetry in the continuum to  $C_3$.  This allows  for  cubic terms in the 
action that are known to lead to ﬁrst order transitions \cite{cubic-terms1,cubic-terms2}. An exception to this rule 
is in the presence of fermions \cite{Li-Z3,Torres18}.
We therefore aim to implement the model on a $\pi$-flux square lattice.  Since the negative sign problem  is avoided due to time 
reversal symmetry doping is accessible.  Hence our model, provides an extremely rich playground for manipulation of topology, and understanding the physics of metallic phases in the proximity of critical points and correlated insulators. \\

The processed data used for generating the figures of this Letter can be accessed under (DOI) 10.58160/t2n8a0h6n21k1mha and the code is available online \cite{code}.

\begin{acknowledgments}
	FFA would like to thank Lukas Janssen for discussions. The authors gratefully acknowledge the Gauss Centre for Supercomputing
	e.V. (www.gauss-centre.eu) for funding this project by providing computing
	time on the GCS Supercomputer SuperMUC-NG at the Leibniz Supercomputing Centre
	(www.lrz.de). Furthermore, the authors acknowledge HPC recources by the Erlangen National High Performance Computing Center (NHR@FAU) of the Friedrich-Alexander-Universität Erlangen-Nürnberg (FAU) under the NHR project b133ae. NHR funding is provided by federal and Bavarian state authorities. NHR@FAU hardware is partially funded by the German Research Foundation (DFG) – 440719683..  FFA and TS thank the W\"urzburg-Dresden
	Cluster of Excellence on Complexity and Topology in Quantum Matter ct.qmat
	(EXC 2147, project-id 390858490) and GR the DFG funded SFB 1170 on Topological
	and Correlated Electronics at Surfaces and Interfaces. MR  acknowledges support by the DFG through Grant No. 332790403.
\end{acknowledgments}

\bibliographystyle{./apsrev4-2}
\bibliography{./literature.bib}

\clearpage

\onecolumngrid
\begin{center}
	\large \bf Supplemental Material for\\
	Manipulating topology of quantum phase transitions by symmetry enhancement
\end{center}
\twocolumngrid

\section{VBS phase and Fierz identity}
\label{sec:Appendix-Fierz-identity}
As the Kekulé order parameter is not directly contained in the model defined in the main text, the understanding of the observed VBS phase is not obvious. The large-$N$ limit does not capture the formation of VBS order, as decomposing and reformulating the interaction term (which is a lattice regularization of the QSH mass term) by employing completeness relations of the Pauli matrices and restructuring the terms into nearest-neighbor interactions does not correlate all bonds of the lattice. Thus, VBS ordering must be dynamically generated involving quantum fluctuations. In order to understand the origin of the Kekulé phase within the notion of mass terms we employ Fierz identity \cite{Igor-Fierz-identity}. A similar principle as discussed in the following has previously been used to circumvent the sign-problem \cite{Ippoliti2018,Assaad-PhysRevLett.126.045701,Meng-Fierz-PhysRevLett.132.246503}.\\
In order to present our explanation of the Kekulé VBS phase, we first derive the specific form of Fierz identity that we employ. Following the definition of the Pauli matrices one can confirm the identity
\begin{eqnarray}
	\frac{1}{2}\textrm{Tr}(\sigma^a\sigma^b)=\delta^{a,b} \quad \textrm{with}\quad a,b=0,1,2,3.
\end{eqnarray}
This can be generalized from $2\times 2$ to $n\times n$ for $n=2^m$, $m\in\mathbb{N}$ by defining the basis of hermitian $n\times n$ matrices as $\Gamma^{a}=\sigma_1^{a_1}\otimes\ldots\otimes\sigma_{m}^{a_{m}}$ with the identity
\begin{eqnarray}
	\frac{1}{2^n}\textrm{Tr}(\Gamma^a\Gamma^b)=\delta^{a,b} \label{eq:Trace-id}
\end{eqnarray}
where $a,b$ now are superindices containing all the indices $a_1,\ldots, a_{m}=0,1,2,3$.
Since any $n\times n$ hermitian matrix $M$ can be written in the form
$M = \sum_{a} x_a \Gamma^a $
the identity 
\begin{align}
	&M = \frac{1}{2^n} \sum_a\textrm{Tr}(M\Gamma^a)\Gamma^a \\
	\Rightarrow &M_{i,j}= \frac{1}{2^n} \sum_a\sum_{n,m}(M_{n,m}\Gamma_{m,n}^a)\Gamma_{i,j}^a \label{eq:Mij}
\end{align}
holds. In equation (\ref{eq:Mij}) we can identify the completeness relation for the basis of hermitian $n\times n$ matrices
\begin{eqnarray}
	\frac{1}{2^n}\sum_a\Gamma_{m,n}^a\Gamma_{i,j}^a= \delta_{n,i}\delta_{m,j} \label{eq:completeness-relation}
\end{eqnarray}
Fierz identity, which can be expressed as
\begin{eqnarray}
	M_{i,j}N_{m,n}=\frac{1}{2^{2n}}\textrm{Tr}(M\Gamma^aN\Gamma^b)\Gamma_{i,n}^b\Gamma_{m,j}^a \label{eq:Fierz-matrices}
\end{eqnarray}
can now be confirmed by using equation (\ref{eq:completeness-relation}) in the right hand side of equation (\ref{eq:Fierz-matrices}):
\begin{align}
	&\textrm{Tr}(M\Gamma^aN\Gamma^b)\Gamma_{i,n}^b\Gamma_{m,j}^a=M_{\alpha,\beta}\Gamma_{\beta,\gamma}^aN_{\gamma,\mu}\Gamma_{\mu,\alpha}^b\Gamma_{i,n}^b\Gamma_{m,j}^a\nonumber\\
	&=M_{\alpha,\beta}N_{\gamma,\mu}2^{2n}\underbrace{\frac{1}{2^n}\Gamma_{\beta,\gamma}^a\Gamma_{m,j}^a}_{\delta_{\gamma,m}\delta_{\beta,j}}\underbrace{\frac{1}{2^n}\Gamma_{\mu,\alpha}^b\Gamma_{i,n}^b}_{\delta_{\alpha,i}\delta_{\mu,n}}
\end{align}
We now want to apply this to mass terms by first considering 
\begin{align}
	&c^\dagger M c c^\dagger N c = c_i^\dagger M_{i,j} c_j c_m^\dagger N_{m,n} c_n \nonumber\\
	&= c_i^\dagger c_j c_m^\dagger c_n \frac{1}{2^{2n}}\textrm{Tr}(M\Gamma^aN\Gamma^b)\Gamma_{i,n}^b\Gamma_{m,j}^a \nonumber\\
	&= \frac{1}{2^{2n}}\textrm{Tr}(M\Gamma^aN\Gamma^b) c_i^\dagger c_j (\delta_{n,m} - c_n c_m^\dagger) \Gamma_{i,n}^b\Gamma_{m,j}^a \nonumber\\
	&= \frac{1}{2^{2n}}\textrm{Tr}(M\Gamma^aN\Gamma^b) \Big(c_i^\dagger \Gamma_{i,n}^b\Gamma_{n,j}^a c_j + c_i^\dagger \Gamma_{i,n}^b c_n c_j \Gamma_{m,j}^a c_m^\dagger\Big) \nonumber
\end{align}
By applying fermionic anticommutation rules in the second term this can be rewritten in the form
\begin{eqnarray}
	(c^\dagger Mc)(c^\dagger Nc) =
	\frac{1}{2^{2n}}\textrm{Tr}(M \Gamma^a  N\Gamma^b)\nonumber\\
	\cdot(c^\dagger\Gamma^b\Gamma^ac - c^\dagger\Gamma^bcc^\dagger\Gamma^ac+c^\dagger\Gamma^bc\textrm{Tr}\Gamma^a) \label{eq:Fierz-id-rewritten}
\end{eqnarray}
As mass terms are represented in terms of $\Gamma^c$, we can now simplify this by setting $M=N=\Gamma^c$ in order to obtain from equation (\ref{eq:Fierz-matrices})
\begin{align}
	\Gamma_{i,j}^c\Gamma_{m,n}^c &= \frac{1}{2^{2n}}\textrm{Tr}(\Gamma^c\Gamma^a\Gamma^c\Gamma^b)\Gamma_{i,n}^b\Gamma_{m,j}^a \nonumber\\
	&= \frac{1}{2^{2n}}f_{c,a}\textrm{Tr}(\Gamma^c\Gamma^c\Gamma^a\Gamma^b)\Gamma_{i,n}^b\Gamma_{m,j}^a\nonumber \\
	&= f_{c,a}\Gamma_{i,n}^a\Gamma_{m,j}^a
\end{align}
where 
\begin{eqnarray}
	f_{c,a}=\frac{1}{2^n}\begin{cases}
		&+1,\quad\textrm{if}\quad \Gamma^c\Gamma^a=\Gamma^a\Gamma^c \\
		&-1,\quad\textrm{if}\quad \Gamma^c\Gamma^a=-\Gamma^a\Gamma^c
	\end{cases}\label{eq:fca}
\end{eqnarray}
and in the last step we used equation (\ref{eq:Trace-id}). Using the same trick in equation (\ref{eq:Fierz-id-rewritten}) yields
\begin{eqnarray}
	(c^\dagger\Gamma^cc)^2 =
	-\sum_a f_{c,a} (c^\dagger \Gamma^a c)^2 + c^\dagger c \Big(\sum_a f_{c,a}+1\Big) \label{eq:Fierz-id-mass-terms}
\end{eqnarray}
which offers a representation of any mass term $\Gamma^c$ in terms of all other possible mass terms $\Gamma^a$ with a sign $f_{c,a}$. In order for a mass term on the right hand side to be energetically favorable it has to have the same sign as the mass term on the left hand side, thus $f_{c,a}$ should be negative. This is the case, only if the mass terms $\Gamma^c$ and $\Gamma^a$ anticommute, i.e. $\{\Gamma^c,\Gamma^a\}=0$.\\

In the notion of masses to the Dirac equation we employ a 16$\times$16 basis of hermitian matrices to consider $\Gamma^a=\nu^{a_1}\sigma^{a_2}\mu^{a_3}\tau^{a_4}$ where $\nu$ denotes the particle-hole (Nambu), $\sigma$ the spin, $\mu$ the valley and $\tau$ the sublattice degree of freedom. Mass terms are then defined as operators $\Gamma^a$ that anticommute with the Dirac Hamiltonian
\begin{eqnarray}
	H = -\frac{\hbar v_F}{2}\sum_{\boldsymbol{k}} \boldsymbol{\psi}^\dagger(\boldsymbol{k})\Big(k_x\mu^z\tau^x-k_y\nu^z\tau^y\Big)\boldsymbol{\psi}(\boldsymbol{k})\label{eq:Dirac-eq}
\end{eqnarray}
where the components of $\boldsymbol{\psi}^\dagger(\boldsymbol{k})$ read
\begin{eqnarray}
	{\psi}_{\nu,\sigma,\mu,\tau}^\dagger(\boldsymbol{k})=\begin{cases}
		&c_{\tau,\mu\boldsymbol{K}+\boldsymbol{k},\sigma}^\dagger\quad\textrm{if}\quad \nu=+1 \\
		&c_{\tau,\mu\boldsymbol{K}-\boldsymbol{k},\sigma}		 \quad\textrm{if}\quad \nu=-1
	\end{cases}
\end{eqnarray}
is a 16-component spinor and $c_{\tau,\mu\boldsymbol{K}+\boldsymbol{k},\sigma}^\dagger$ creates an electron with spin $\sigma$ and momentum $\boldsymbol{k}$ in valley $\mu$ and sublattice $\tau$ of the honeycomb lattice \cite{Toshihiro-mass-terms}. Starting from the more widely used form of the Dirac Hamiltonian
\begin{align}
	H = -\frac{\hbar v_F}{2}\sum_{\boldsymbol{k}} \boldsymbol{\phi}^\dagger(\boldsymbol{k})\Big(k_x\mu^z\tau^x-k_y\tau^y\Big)\boldsymbol{\phi}(\boldsymbol{k})\label{eq:Dirac-eq-8}
\end{align}
with 8-component spinors $\boldsymbol{\phi}^\dagger(\boldsymbol{k})=(c_{\tau,\mu\boldsymbol{K}+\boldsymbol{k},\sigma}^\dagger)$, the Hamiltonian in equation (\ref{eq:Dirac-eq}) is obtained by transforming into a Bogoliubov basis using the Bogoliubov-de-Gennes Hamiltonian
\begin{align}
	H_{\textrm{BdG}} = \sum_{\boldsymbol{k}}\boldsymbol{\psi}^\dagger(\boldsymbol{k})\begin{pmatrix}
		\varepsilon(\boldsymbol{k}) & \Delta(\boldsymbol{k}) \\
		-\Delta^*(-\boldsymbol{k}) & -\varepsilon^*(-\boldsymbol{k})
	\end{pmatrix}\boldsymbol{\psi}(\boldsymbol{k})
\end{align}
By symmetry considerations we can identify different mass terms to the Dirac equation. All mass-terms have to fulfill particle-hole symmetry (PHS) which can be defined by the symmetry transformation
\begin{eqnarray}
	P X(k) P^{-1} = -X(-k) \quad \textrm{with} \quad P = \nu^xK
\end{eqnarray}
where $K$ denotes complex conjugation and $X(k)$ is any operator that is invariant under the given symmetry transformation \cite{Mudry2009}. Other symmetry conditions reflect time reversal symmetry (TRS)
\begin{align}
	TX(-k)T^{-1}=X(k) \quad \textrm{with}\quad T=i\sigma^y\mu^x K,
\end{align}
sublattice (= chiral) symmetry (SLS)
\begin{align}
	SX(k)S^{-1}=-X(k) \quad \textrm{with}\quad S=\tau^z,
\end{align}
as well as SU(2) spin rotation symmetry (SRS)
\begin{align}
	UX(k)U^{-1} = X(k)\quad \textrm{with}\quad U=\begin{pmatrix}
		e^{i\theta\frac{\hbar}{2}\boldsymbol{\sigma}\cdot\boldsymbol{n}} & 0 \\
		0 & e^{-i\theta\frac{\hbar}{2}\boldsymbol{\sigma}^*\cdot\boldsymbol{n}}
	\end{pmatrix}
\end{align}
with  $U$ accounting for the structure of the BdG Hamiltonian. It follows that SRS is fulfilled if
\begin{align}
	[\nu^z\sigma^x,X]=[\sigma^y,X]=[\nu^z\sigma^z,X]=0
\end{align}
Note that the Dirac Hamiltonian (\ref{eq:Dirac-eq}) preserves all the above symmetries. Mass terms can now be classified by different symmetry breakings. All mass-terms have to anticommute with the Dirac Hamiltonian and preserve PHS.
Reflecting the fact that the three QSH mass terms anticommute with the Hamiltonian in equation (\ref{eq:Dirac-eq}), break SLS and SRS but preserve TRS, they can be identified as
\begin{align}
	&\Gamma_{\textrm{QSH}}^x=\nu^{z}\sigma^{x}\mu^{z}\tau^{z}, \\
	&\Gamma_{\textrm{QSH}}^y=\hphantom{\nu^0}\sigma^{y}\mu^{z}\tau^{z}, \\
	\textrm{and}\hspace{10pt} &\Gamma_{\textrm{QSH}}^z=\nu^{z}\sigma^{z}\mu^{z}\tau^{z},
\end{align}
whereas the VBS mass terms are given by $\textrm{Re}(\Gamma_{\textrm{VBS}})=\nu^{z}\mu^{x}\tau^{x}$ and $\textrm{Im}(\Gamma_{\textrm{VBS}})=\mu^{y}\tau^{x}$, thus preserving all the above symmetries. It can quickly be confirmed that none of the QSH mass terms anticommute with the VBS mass terms. In fact, they commute with each other yielding a positive sign in $f_{c,a}$ as becomes clear from equation (\ref{eq:fca}). This means that VBS order is not allowed from that point of view. (A full overview over mass terms to the Dirac equation and tuples of mutually anticommuting mass terms is presented in Ref. \cite{Mudry2009} where the authors work in a chiral representation of the spinors. This leads to a different form of the mass terms than in our consideration but does not influence their commutation relations.)\\
Nevertheless, one can show that the interaction Hamiltonian of our model can be rewritten in the form (for simplicity at $N=1$)
\begin{align}
	\hat{H}_{\lambda}=&-\lambda\sum_{\varhexagon}\sum_{\langle\langle\boldsymbol{i},\boldsymbol{j}\rangle\rangle\in \varhexagon}\sum_{\langle\langle\boldsymbol{i}',\boldsymbol{j}'\rangle\rangle\neq\langle\langle\boldsymbol{i},\boldsymbol{j}\rangle\rangle} \hat{\boldsymbol{J}}_{\langle\langle\boldsymbol{i},\boldsymbol{j}\rangle\rangle}\cdot\hat{\boldsymbol{J}}_{\langle\langle\boldsymbol{i}',\boldsymbol{j}'\rangle\rangle} \label{eq:model-rewritten}\\
	&-\lambda\sum_{\varhexagon}\sum_{\langle\langle\boldsymbol{i},\boldsymbol{j}\rangle\rangle\in \varhexagon}\Big(6\hat{\eta}_{\boldsymbol{i}}^\dagger\hat{\eta}_{\boldsymbol{j}}+\textrm{h.c.}-4\hat{\boldsymbol{S}}_{\boldsymbol{i}}\cdot\hat{\boldsymbol{S}}_{\boldsymbol{j}}\nonumber\\
	&-5\hat{n}_i\hat{n}_j+5(\hat{n}_i+\hat{n}_j)\Big)\nonumber
\end{align}
where
\begin{align}
	&\hat{\boldsymbol{J}}_{\langle\langle\boldsymbol{i},\boldsymbol{j}\rangle\rangle}=i\nu_{\boldsymbol{i}\boldsymbol{j}}\hat{\boldsymbol{c}}_{\boldsymbol{i}}^\dagger\boldsymbol{\sigma}\hat{\boldsymbol{c}}_{\boldsymbol{j}}+\textrm{h.c.},\nonumber\\
	&\hat{\eta}_{\boldsymbol{i}}^\dagger=\hat{c}_{\boldsymbol{i},\uparrow}^\dagger\hat{c}_{\boldsymbol{i},\downarrow}^\dagger,\quad \hat{\eta}_{\boldsymbol{i}}=\hat{c}_{\boldsymbol{i},\downarrow}\hat{c}_{\boldsymbol{i},\uparrow}, \\
	&\hat{\boldsymbol{S}}_{\boldsymbol{i}}=\frac{1}{2}\hat{\boldsymbol{c}}_{\boldsymbol{i}}^\dagger\boldsymbol{\sigma}\hat{\boldsymbol{c}}_{\boldsymbol{j}}.\nonumber
\end{align}
as stated in the supplemental material of \cite{Zhenjiu-QSH-SC}. In this formulation it becomes visible, that there is an attractive pairing term contained in the Hamiltonian, which gives rise to the observed SSC phase, breaking the U(1) charge symmetry of the Hamiltonian. It allows for a mean field term
\begin{eqnarray}
	H_{\textrm{SSC}} = -\lambda\phi \sum_{\boldsymbol{i}}(c_{\boldsymbol{i},\uparrow}^\dagger c_{\boldsymbol{i},\downarrow}^\dagger + \textrm{h.c.}) 
\end{eqnarray}
which opens a mass gap that we can account for by considering SSC mass terms to the Dirac equation. Those are given by $\textrm{Re}(\Gamma_{\textrm{SSC}})=\nu^{y}\sigma^{y}\mu^{x}$ (breaking SLS) and $\textrm{Im}(\Gamma_{\textrm{SSC}})=\nu^{x}\sigma^{y}\mu^{x}$ (breaking SLS and TRS) \cite{Toshihiro-mass-terms, Mudry2009}. These mass terms both anticommute with the VBS mass terms, which results in a negative sign for $f_{c,a}$ in equation (\ref{eq:fca}). This suggests that the Fierz identity in equation (\ref{eq:Fierz-id-mass-terms}) may allow for VBS ordering in the presence of the SSC mass terms. This could explain why VBS order is observed only in the regime where the SSC phase is energetically favorable compared to the QSH phase, indicating that the SSC phase might support or stabilize VBS order. In this scenario, the SSC phase can be seen as facilitating the emergence of VBS order. Furthermore, in the large-$N$ limit, where the SSC phase disappears, VBS order is also not observed. This is because the second term in equation (\ref{eq:model-rewritten}) acquires a factor of $1/N$, making the first term dominant. In the $N=1$ case, VBS order does not emerge. One possible interpretation is that, for $N=1$, the SSC phase involves the condensation of Skyrmions, and it may require too much energy to dissolve them, preventing the formation of VBS order. In the $N=2$ case, the absence of such topological effects might allow the system to stabilize VBS order as a lower-energy configuration.\\

\section{Mean field}
\label{sec:mean_field}
\begin{figure}[htb]
	\centering
	\includegraphics[width=\columnwidth]{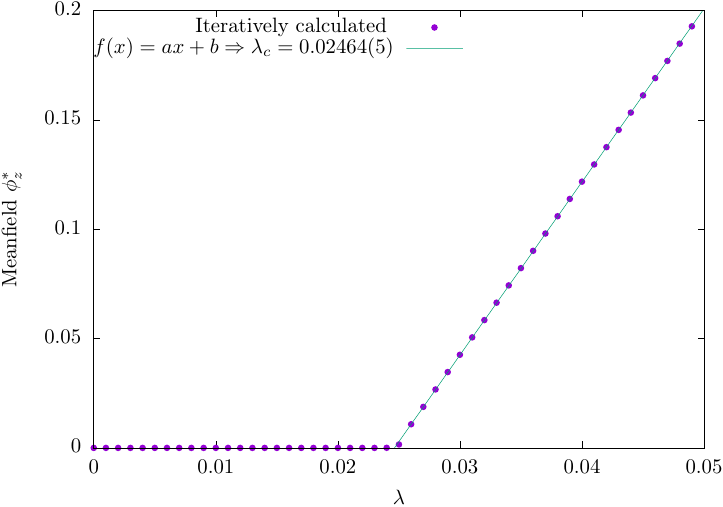}
	\caption{Hubbard-Stratonovich field $\phi_z^*$ calculated by self-consistently solving the saddle point equation for different values of $\lambda$. Fitting the data with a linear function reveals a transition point between SM and QSH at $\lambda_c=0.02464(5)$.}
	\label{fig:Meanfield}
\end{figure}
A self-consistent calculation of the mean field by neglecting spatial and temporal fluctuations $\phi_{i,\tau}^z\rightarrow\phi_z^*$ reveals that only an SM and a QSH phase are observed as $N\to\infty$. It is straightforward to perform a Hubbard-Stratonovich transformation in the partition sum of the model in the main text in order to derive the saddle point equation
\begin{align}
	\boldsymbol{\phi}_{n,\tau}^* = 2\lambda \big\langle\hat{\boldsymbol{O}}_{n}\big\rangle_{\boldsymbol{\phi}_{n,\tau}^*}\label{eq:self-consistency}
\end{align}
where $\hat{\boldsymbol{O}}_{n}=\sum_{\ll\boldsymbol{i},\boldsymbol{j}\gg\in n}i\nu_{\boldsymbol{i}\boldsymbol{j}}\boldsymbol{c}_{\boldsymbol{i}}^\dagger\boldsymbol{\sigma}\boldsymbol{c}_{\boldsymbol{j}}+\text{h.c.}$ from the action, which is independent on the SU($N$) flavor degree of freedom due to the block structure of the matrix. This, after neglecting spatial and temporal fluctuations leads to 
\begin{align}
	\boldsymbol{\phi}^*
	=
	2\lambda \frac{\text{Tr}\left[e^{-\beta \hat{H}_{MF}} \hat{\boldsymbol{O}}_{n}\right]}{\text{Tr}\left[e^{-\beta\hat{H}_{MF}}\right]}
\end{align}
with 
\begin{align}
	\hat{H}_{MF}=&H_t-\boldsymbol{\phi}^*\sum_n\boldsymbol{O}_n
\end{align}
Fourier transformation yields the expression
\begin{align}
	\phi_z^* =& -4\frac{\lambda}{N_{\varhexagon}} \sum_{\boldsymbol{k},\sigma}\text{Im}\left[M(\boldsymbol{k})\right]\sigma\Big\langle\left(a_{\boldsymbol{k},\sigma}^\dagger a_{\boldsymbol{k},\sigma}-b_{\boldsymbol{k},\sigma}^\dagger b_{\boldsymbol{k},\sigma}\right)\Big\rangle_{\boldsymbol{\phi}^*}
	\label{eq:self-consistency2}
\end{align}
with $M(\boldsymbol{k}) =e^{i\boldsymbol{k}\cdot\boldsymbol{a}_1}+e^{-i\boldsymbol{k}\cdot\boldsymbol{a}_2}+e^{-i\boldsymbol{k}\cdot(\boldsymbol{a}_1-\boldsymbol{a}_2)}$, $\sigma=\pm 1$ and $N_{\varhexagon}$ the number of hexagons. We can iteratively solve this equation by evaluating the expectation value
\begin{align}
	\Big\langle a_{\boldsymbol{k},\sigma}^\dagger a_{\boldsymbol{k},\sigma}\Big\rangle_{\boldsymbol{\phi}^*}= \sum_{n=1}^2 \left|(U_{\boldsymbol{k},\sigma})_{1,n}\right|^2 f(E_n(\boldsymbol{k}))
\end{align}
for an initial $\phi_z^*$ where the unitary matrix $U_{\boldsymbol{k},\sigma}$ diagonalizes the Fourier transformed mean field Hamiltonian leading to the spectrum $E_n(\boldsymbol{k})$. A mean-field analysis for different values of $\lambda$ results in a transition point between SM and QSH at $\lambda_c=0.02464(5)$ as shown in Fig. \ref{fig:Meanfield} which is not much greater than the transition point that we observe in QMC.

\section{Additional data}
\begin{figure*}[htb]
	\centering
	\includegraphics[width=0.195\linewidth]{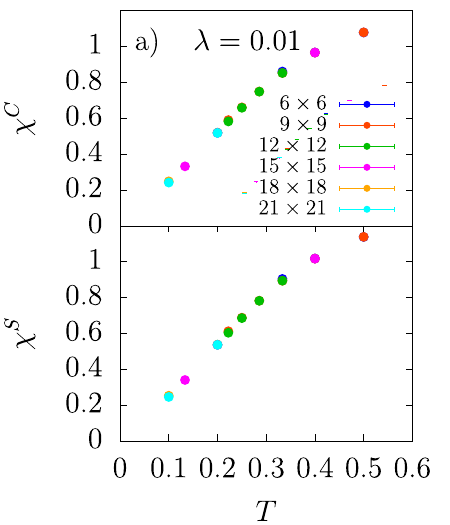}
	\hspace{-.2cm}
	\includegraphics[width=0.195\linewidth]{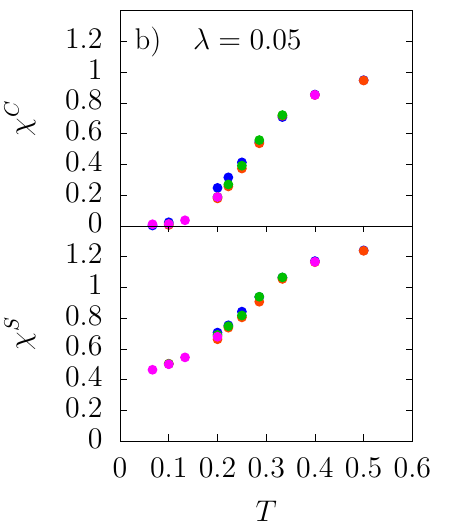}
	\hspace{-.2cm}
	\includegraphics[width=0.195\linewidth]{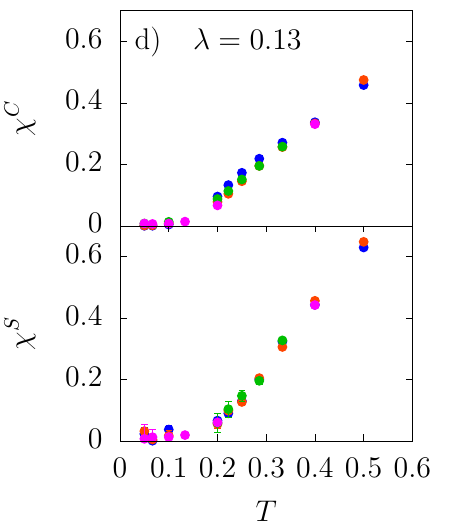}
	\hspace{-.2cm}
	\includegraphics[width=0.195\linewidth]{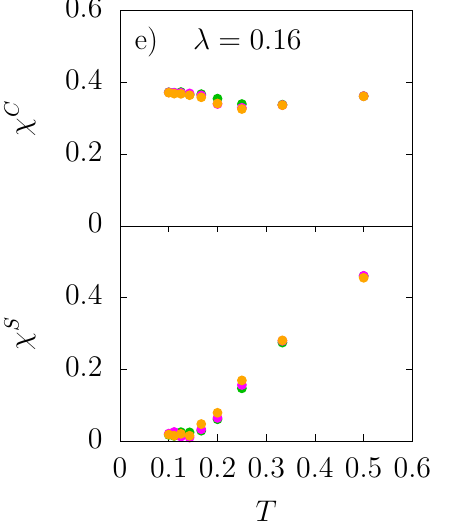}
	\hspace{-.2cm}
	\includegraphics[width=0.195\linewidth]{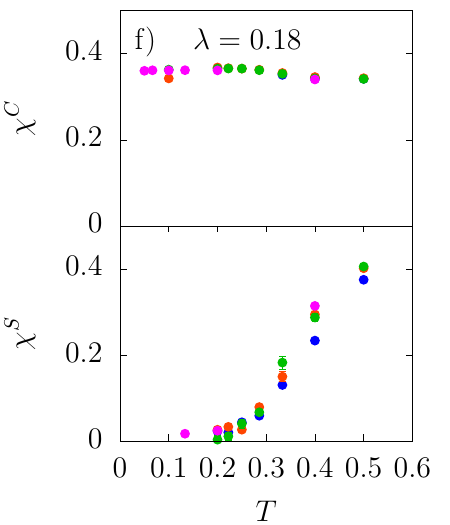}
	\caption{Uniform charge (top) and spin (bottom) susceptibilities $\chi^{C/S}$ of the (a) semimetal, (b) QSH, (d) Kekulé and (f) SSC phase as well as at the 
		(e) Kekulé-SSC transition versus temperature $T$.}
	\label{fig:uniform-sus}
\end{figure*}
\begin{figure}[htb]
	\centering
	\includegraphics[width=\columnwidth]{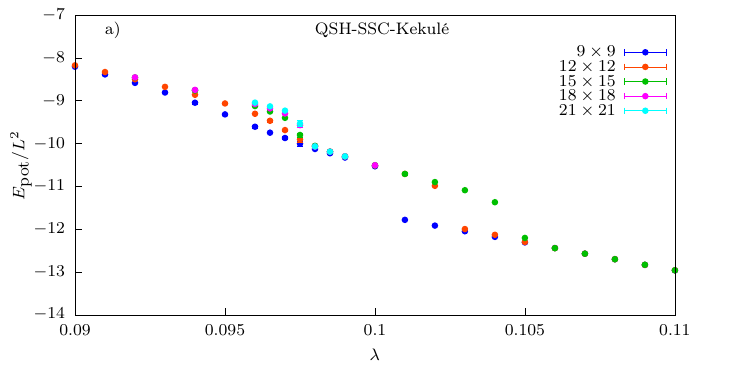}
	\caption{Potential energy as a function of $\lambda$. The QSH-SSC transition shows a jump for larger lattice sizes while the SSC-Kekule transition shows a clearly defined jump also for small lattice sizes. The position of this jump is very sensitive to the lattice size.}
	\label{fig:pot_scal_QSH-SSC}
\end{figure}
In order to complement the picture that we gave in the main text, we present some additional data. Figure \ref{fig:uniform-sus} shows charge and spin susceptibilities in the different phases. The semimetal phase shows a linear dependence of the susceptibilities as $T\to 0$ which was already mentioned in the main text (Fig. \ref{fig:uniform-sus}(a)). In the QSH phase a charge gap opens whereas spin excitations are gapless (Fig. \ref{fig:uniform-sus}(b)). This gives rise to the formation of Skyrmions. In the Kekulé VBS phase both charge and spin excitations are gapped which is in accordance to the fact that this is a trivial insulator (Fig. \ref{fig:uniform-sus}(c)). In the coexistence as well as the pure SSC phase we observe a spin gap while charge excitations are gapless which is necessary for the transport of charge carriers (Fig. \ref{fig:uniform-sus}(d) and Fig. \ref{fig:uniform-sus}(e)).\\

In order to support the first order behavior of the QSH-SSC and the SSC-VBS transition, we plot the scaled potential energy $E_{\textrm{pot}}=-\sum_n\langle\hat{O}_n^2\rangle$ of the system in Fig. \ref{fig:pot_scal_QSH-SSC}. In the case of a first order transition, the free energy of the system is supposed to show a kink as a function of $\lambda$. In the ground state ($T=0$) the free energy equals the internal energy of the system. Therefore, stemming from the shape of the Hamiltonian, the potential energy can be identified with the derivative of the free energy with respect to $\lambda$. As such, it is expected to indicate a jump in the case of a first order transition. Whereas the jump in the SSC-VBS transition is very clearly visible in all of the shown lattice sizes, the jump in the QSH-SSC transition is only noticeable in the lattices with $L=18$ and $L=21$. However, it overall fits their categorization as first order transitions.\\
\begin{figure}[htb]
	\centering
	\includegraphics[width=\columnwidth]{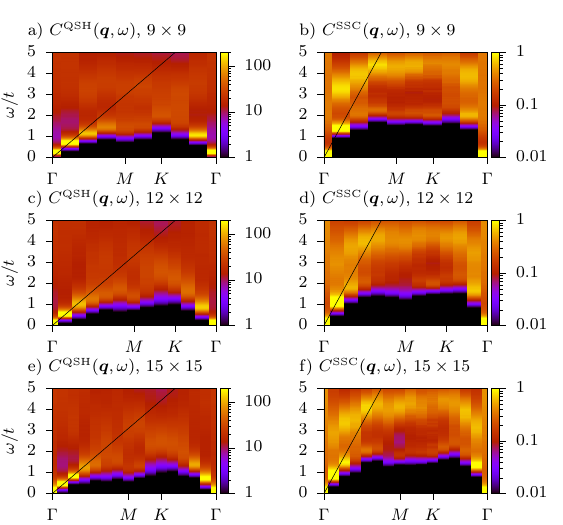}
	\caption{Dynamical QSH (a,c,e) and s-wave pairing (b,d,f) structure factors at $\lambda=0.0975$ for different lattice sizes $L=9,12,15$. The slope denotes the respective velocities  which are consistent across different lattice sizes.}
	\label{fig:QSH-SSC-spectral-crit}
\end{figure}
\begin{figure}[htb]
	\centering
	\includegraphics[width=\columnwidth]{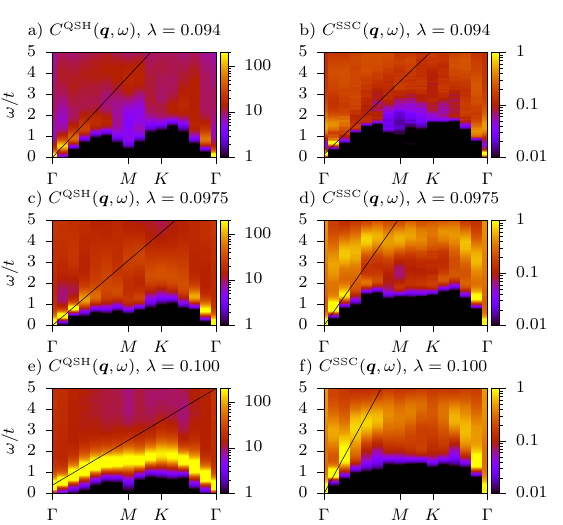}
	\caption{Dynamical QSH (a,c,e) and s-wave pairing (b,d,f) structure factors at different values of $\lambda$ for $L=15$. The slope denotes the respective velocities.}
	\label{fig:QSH-SSC-spectral}
\end{figure}

In 
the main text, it was shown that the Fermi velocity, extracted from the dynamical structure factors differs by a factor of 2, reflecting the fact that Lorentz symmetry is broken. This is supported here, by presenting the dynamical structure factors for different lattice sizes in Fig. \ref{fig:QSH-SSC-spectral-crit}. They are obtained by calculating time displaced correlation functions and deconvolute the data by a stochastic analytical continuation approach, included in the ALF package as a maximum entropy algorithm. Since the difference in velocity is stable across all the considered lattice sizes, this shows that it is not a finite size effect. Moreover, in Fig. \ref{fig:QSH-SSC-spectral} we plot the dynamical structure factors for QSH and SSC correlations at different values of $\lambda$ around the QSH-SSC transition. We resolve the gap in the respective order parameter before and after the transition and show that the velocity continuously shifts with $\lambda$, decreasing for QSH while increasing for SSC correlations with increasing $\lambda$. At the point, where the velocities align, the SSC structure factor still shows a small gap.

\subsection{Histograms}
\begin{figure}[htb]
	\centering
	\includegraphics[width=0.32\linewidth]{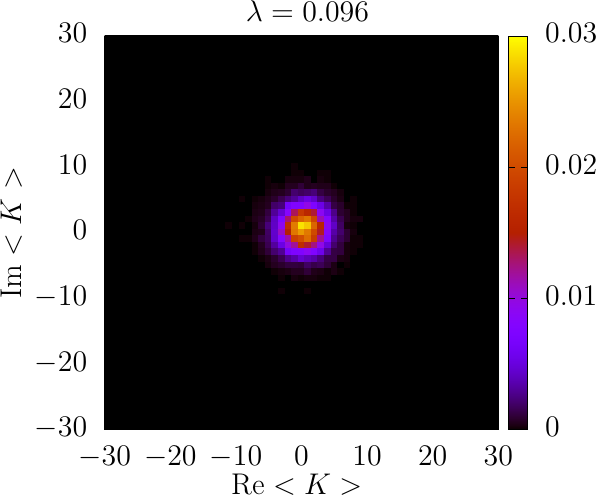}
	\hfill
	\includegraphics[width=0.32\linewidth]{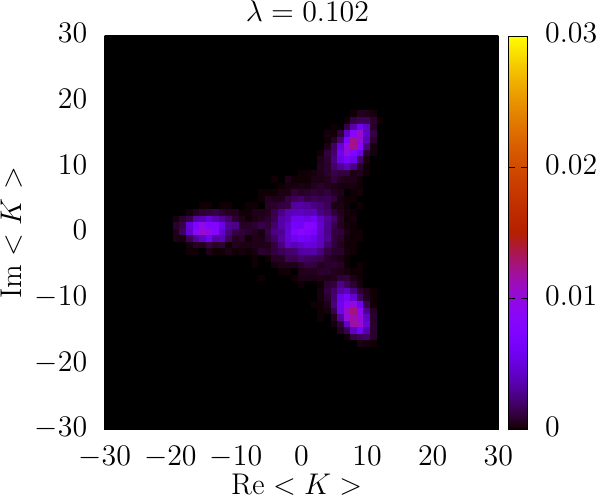}
	\hfill
	\includegraphics[width=0.32\linewidth]{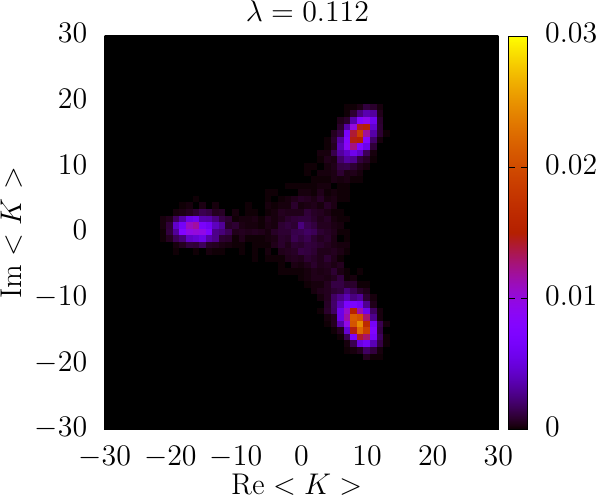}
	\includegraphics[width=0.32\linewidth]{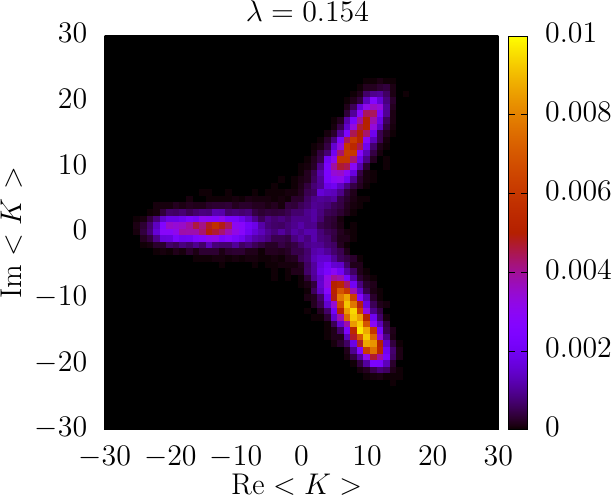}
	\hfill
	\includegraphics[width=0.32\linewidth]{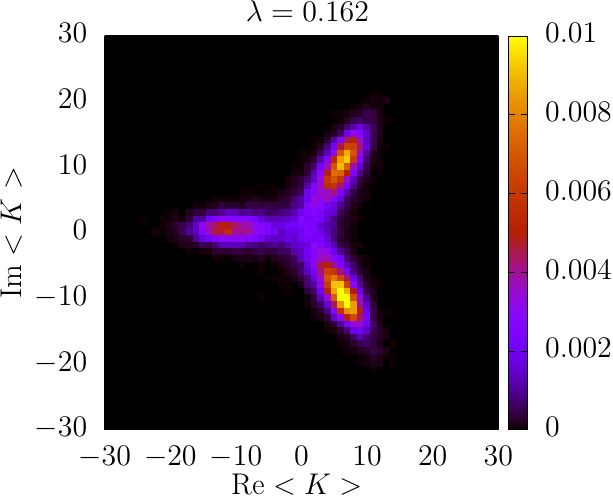}
	\hfill
	\includegraphics[width=0.32\linewidth]{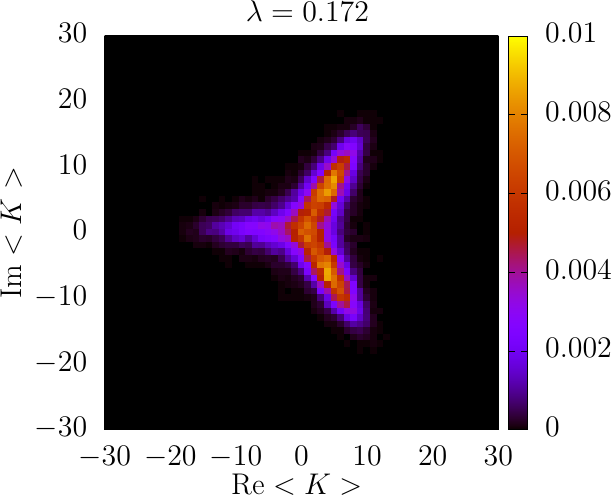}
	\caption{Histograms for the Kekulé order parameter around the SSC-Kekulé transition (top) and around the coexistence region (bottom) on the $6\times6$ lattice for $\beta=25$ and $\Delta\tau=0.05$.}
	\label{fig:hist-SSC-Kekule}
\end{figure}

The nature of phase transitions can also be explored by plotting histograms of various observables around the transition points. These histograms represent probability distributions of specific order parameters. After a set number of sweeps (i.e., each bin), a QMC snapshot is taken by measuring the observables and sorting their values into a discrete grid. The data points thus represent counts of measurement events that result in a value of the observables which falls into a given interval. They are normalized with respect to the total number of counts. In order to generate a histogram for the Kekulé observable, we use the order parameter
\begin{align}
	K =&  \frac{1}{\sqrt{V}} \sum_{\boldsymbol{r}} \left(K_{\boldsymbol{r},1} + e^{i\frac{2\pi}{3}}K_{\boldsymbol{r},2} + e^{i\frac{4\pi}{3}}K_{\boldsymbol{r},3}\right)e^{i\boldsymbol{K}\cdot\boldsymbol{r}} \\
	&\text{with}\quad K_{\boldsymbol{r},\mu} = \sum_{\sigma,s} c_{\boldsymbol{r},s,\sigma}^\dagger c_{\boldsymbol{r}+\boldsymbol{\delta}_\mu,s,\sigma} + \text{h.c.} \nonumber
\end{align}
where the values belonging to different bonds of the honeycomb lattice are split into three different groups $\mu=1,2,3$ in the complex plane. Here, $\boldsymbol{\delta}_{\mu}$ defines the position of the respective nearest neighbor of the orbital at position $\boldsymbol{r}$. The histograms of the Kekulé order parameter are shown in Fig. \ref{fig:hist-SSC-Kekule} for different values of $\lambda$ around the SSC-Kekulé transitions. The histograms in the top row are compatible with our findings, that this is a first order transition, where one ordered state is abruptly replaced by another. Around the coexistence region (bottom row of Fig. \ref{fig:hist-SSC-Kekule}) we observe that the three peaks, belonging to the different bonds, become elongated towards the center while slowly merging into each other until the system cannot resolve them anymore. For a quantum critical point, we would expect for the Kekulé order parameter to reveal a ring-shaped distribution at the critical point, reflecting the enhanced SO(4) symmetry that emerges, locking together the order parameters of SSC and Kekulé into a four-component order parameter. This cannot be observed here.\\
It may be interesting to note, that the weighting of the different peaks depends on the imaginary time discretization $\Delta\tau$, since it breaks symmetry. Thus, $\Delta\tau$ has to be chosen sufficiently small in order to observe a balanced weighting among the different bonds.

\subsection{Coexistence between Kekulé-VBS and SSC}
\begin{figure}
	\centering
	\includegraphics[width=\columnwidth]{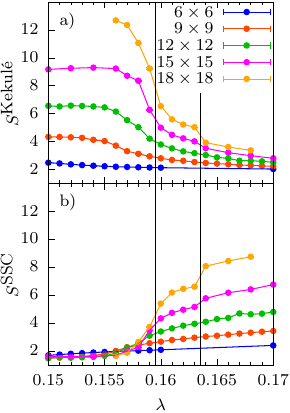}
	\caption{Structure factors of (a) Kekulé order at the Dirac-point and (b) SSC order at the $\Gamma$-point as a function of $\lambda$ around the Kekulé-SSC transition. The vertical bars show a coexistence region where both structure factors diverge with increasing lattice size.}
	\label{fig:coexistence}
\end{figure}
\begin{figure}[htb]
	\centering
	\includegraphics[width=\columnwidth]{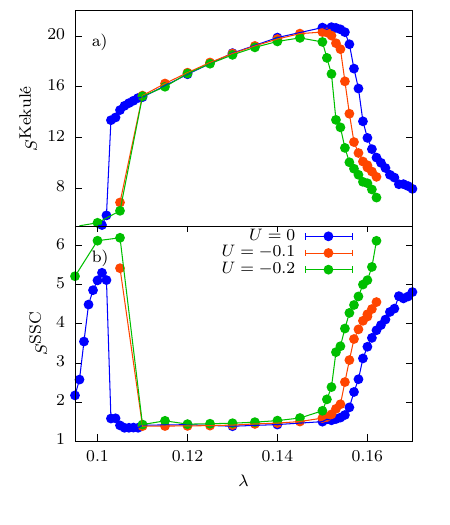}
	\caption{Structure factors of (a) Kekulé order at the Dirac-point and (b) SSC order at the $\Gamma$-point at $L=12$ for different attractive Hubbard interactions.}
	\label{fig:Kekule-Hubbard}
\end{figure}
In the main text, we have shown that at $\lambda=0.16$ both long-range Kekulé-VBS and SSC order are present. In fact, this phase coexistence is extended over a particular range, as marked by the two vertical bars in Fig. \ref{fig:coexistence}. Within this range both the Kekulé and SSC structure factors diverge with increasing lattice size.\\
Generically, to probe phase coexistence we tried to exploit the fact, that the two transitions between ordered and disordered phase are occurring separately for the two involved order parameters. In contrast, a quantum critical point would lock the two order parameters by an enhanced symmetry at the critical point, thus yielding a single transition of one ordered phase into the other.\\
A first idea relies on changing the chemistry of the system by tuning the imaginary time discretization $\Delta\tau$ and thus the energy scales of the system. Following the above explanation, the size of the coexistence region should depend on that. Since in this $\lambda$ range the system suffers from long auto-correlation times, it was difficult to reduce $\Delta\tau$.\\
Another way to analyse the coexistence phase is to complement the Hamiltonian with an additional attractive Hubbard-$U$ term. As Fig. \ref{fig:Kekule-Hubbard} reveals, increasing $U$ leads to the suppression of Kekulé-VBS order. Close to the coexistence region, the structure factor of Kekulé order decreases while the structure factor of SSC order increases with increasing $U$ leading to a shift of the region to lower $\lambda$. It is expected, that for large values of $U$, the Kekulé-VBS phase will completely vanish, since SSC order dominates over Kekulé-VBS order due to the larger weight of attractive pairing terms in the model. Nevertheless, we did not observe a significant change in the size of the coexistence region.\\
For the coexistence phase to remain thermally stable the particle number $\langle\hat{N}\rangle$ is supposed to increase as a function of chemical potential $\mu$. This can be verified by considering the identity
\begin{equation}
	\frac{\partial}{\partial\mu}\langle\hat{N}\rangle = \beta \left(\langle\hat{N}^2\rangle-\langle\hat{N}\rangle^2\right)
\end{equation}
where the right hand side can be linked to the uniform charge susceptibility. As can be seen in Fig. \ref{fig:uniform-sus}(e) this is finite in the thermodynamic limit at $\lambda=0.16$ hence corresponding to a non-zero derivative of the particle number with respect to chemical potential. It can be argued to call this phase a supersolid \cite{Onsager-supersolid,Andreev-1971,Chester-PhysRevA.2.256,Legget-PhysRevLett.25.1543,Batrouni-PhysRevLett.74.2527,Batrouni-PhysRevLett.84.1599,Fisher-supersolid,Fisher-PhysRevLett.32.1350,Frey-PhysRevB.49.9723}. Yet, it has to be noted that the Kekulé phase only reflects bond order without displaying charge modulation in the sense of density order.

\subsection{Unbroken SU($N$) flavor symmetry}
\begin{figure}[htb]
	\centering
	\includegraphics[width=\columnwidth]{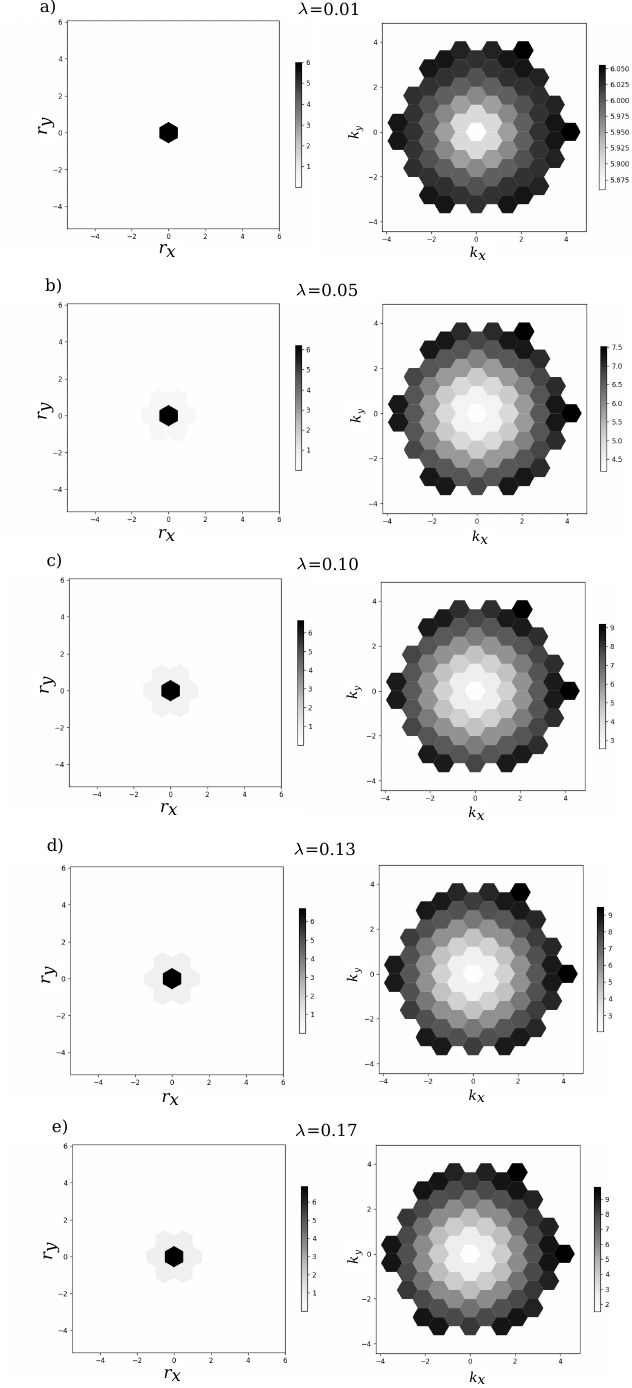}
	\caption{Absolute value of real space correlation functions (left) and structure factors $S(\boldsymbol{k})$ (right) of the generators of SU(2) flavor symmetry for different values of $\lambda$ and $L=9$.}
	\label{fig:SU2col}
\end{figure}

At $N=2$ the generators of SU($N$) flavor symmetry are given by
\begin{equation}
	\hat{T}_{\ve{i}}^{\alpha}= \sum_{\sigma,s,s'}\hat{c}^{\dagger}_{\ve{i},\sigma,s}\tau_{s,s'}^{\alpha}\hat{c}_{\ve{i},\sigma,s'}.
\end{equation}
Here, $\tau^{\alpha}$ is a Pauli matrix in the basis of fermionic flavor $s$. This can be used as an operator for measuring correlation functions in order to compute the structure factor $S(\boldsymbol{k})$ as done for the QSH, SSC and Kekulé VBS order parameters in the main text. In Fig. \ref{fig:SU2col} the real-space correlation function and the structure factor are illustrated for different values of $\lambda$. As obvious from the rapid decline of the correlation function down to zero and a smooth distribution of the structure factor in momentum space, it does not show order in any of the observed phases. Thereby, we argue that SU($N$) flavor symmetry is not broken in the considered SU(2) case.

\end{document}